\documentclass[aps,showpacs,nofootinbib]{revtex4}

\usepackage{graphicx}
\usepackage{graphics}
\usepackage{amssymb}
\usepackage{amsmath}

\newcommand{\half}{\textstyle{\frac{1}{2}}}

\newcommand{\nslash}{\kern 0.2 em n\kern -0.50em /}
\newcommand{\kslash}{\kern 0.2 em k\kern -0.45em /}
\newcommand{\pslash}{\kern 0.2 em p\kern -0.50em /}
\newcommand{\Sslash}{\kern 0.2 em S\kern -0.50em /}
\newcommand{\Pslash}{\kern 0.2 em P\kern -0.50em /}
\newcommand{\Rslash}{\kern 0.2 em R\kern -0.50em /}
\newcommand{\open}{{<\kern -0.3 em{\scriptscriptstyle )}}}

\newcommand{\de}{d}
\newcommand{\st}{T}

\newcommand{\rrho}{\frac{|{\bf{P}}_{C\perp}|}{\sqrt{s}}}
\newcommand{\Gangle[1]}{\delta \hat{G}^{\open #1}}

\begin{document}
\title{
Dihadron interference fragmentation functions in proton-proton collisions
}

\author{Alessandro Bacchetta}
\email{alessandro.bacchetta@physik.uni-regensburg.de}
\affiliation{Institut f{\"u}r Theoretische Physik, Universit{\"a}t Regensburg,
D-93040 Regensburg, Germany}

\author{Marco Radici}
\email{radici@pv.infn.it}
\affiliation{Dipartimento di Fisica Nucleare e Teorica, Universit\`{a} di Pavia, and\\
Istituto Nazionale di Fisica Nucleare, Sezione di Pavia, I-27100 Pavia, Italy}

\begin{abstract}
We study the production of hadron pairs in proton-proton collisions, 
selecting pairs with large total transverse momentum with respect to 
the beam, and small relative transverse momentum, 
i.e., belonging to a single jet with large transverse momentum. 
We describe the process in terms of dihadron fragmentation functions.
We consider the production of one pair in polarized collisions (with one transversely
polarized proton) and the production of two pairs in
unpolarized collisions.
In the first case, we discuss how to observe the quark transversity distribution 
in connection with a specific class of dihadron fragmentation functions, named 
interference fragmentation functions. In the second case, we suggest how to determine 
the latter and also how to observe linearly polarized gluons.
\end{abstract}


\maketitle

\section{Introduction}
\label{sec:intro}

Spin measurements in high-energy collisions of hadrons can be a powerful and 
versatile way to investigate the dynamics of quarks and gluons. For instance, 
several measurements of Single-Spin Asymmetries (SSA) with polarized protons have 
delivered unexpected results (like the observation of large $A_N$ asymmetries by 
the FNAL E704~\cite{Adams:1991cs} and, more recently, by the STAR~\cite{Adams:2003fx} 
collaborations), 
which cannot be justified in the context of perturbative QCD at the partonic 
level~\cite{Kane:1978nd}. Unless invoking subleading-twist 
effects~\cite{Qiu:1991pp,Kanazawa:2000hz}, these large asymmetries can be explained by 
allowing partons both to have an intrinsic transverse 
momentum~\cite{Anselmino:1995tv} and to undergo final-state
interactions 
(which prevents the application of constraints by time-reversal symmetry, leading 
to the so-called T-odd distribution and fragmentation functions). This interpretation has 
been further strengthened by recent measurements of SSA with lepton beams by the 
HERMES~\cite{Airapetian:2004tw,Airapetian:2000tv,Airapetian:2001eg,Airapetian:2002mf} and the
SMC~\cite{Bravar:2000ti} collaborations, as well as of beam spin asymmetries by the CLAS
collaboration~\cite{Avakian:2003pk}. Remarkably, these results can be interpreted 
as effects due to the orbital angular momentum of partons inside the parent
hadron~\cite{Brodsky:2002cx,Belitsky:2002sm,Burkardt:2003je}.

SSA are important not only to study T-odd mechanisms, but also because they can be used 
as analyzing powers of the quark spin and they allow the measurement of quantities 
otherwise inaccessible. The most renowned example is the transversity distribution 
$h_1$~\cite{Ralston:1979ys}, a leading-twist parton density that describes the distribution 
of transversely polarized quarks inside transversely polarized hadrons, an 
essential piece of information to describe the partonic spin structure of 
hadrons~\cite{Jaffe:1991kp,Barone:2003fy}. Being related to a helicity flipping 
mechanism (since helicity and chirality coincide at leading twist, it is usually named as a chiral-odd distribution), it is
suppressed in inclusive Deep-Inelastic Scattering (DIS) because it 
requires another chiral-odd partner.

The idea of accessing transversity at leading twist in nucleon-nucleon collisions has been 
extensively discussed in the literature. The simplest option is Drell-Yan
leptoproduction 
with two transversely 
polarized protons~\cite{Ralston:1979ys,Artru:1990zv,Jaffe:1991kp,Cortes:1992ja,Ji:1992ev},
where the needed chiral-odd partner is the transversity distribution of 
the corresponding antiquark. 
This option seems not promising at operating 
experimental facilities, because the probability of having a transversely 
polarized antiquark in a transversely polarized proton is 
suppressed~\cite{Martin:1998rz,Barone:1997mj}. The situation should be more 
favorable in the future High Energy Storage Ring at GSI (GSI 
HESR)~\cite{Efremov:2004qs,Anselmino:2004ki,pax,Radici:2004ij}, where (polarized) 
antiprotons will be produced. Another option is to consider a semi-inclusive process. For
example, if transversely polarized hyperons are produced in collisions where one of the 
protons is transversely polarized, $h_1$ appears in connection with the fragmentation 
function $H_1$~\cite{deFlorian:1998ba,deFlorian:1998am}. Alternatively, when a single 
unpolarized hadron (e.g., a pion) is inclusively produced in hadronic collisions, $h_1$ 
is convoluted with the chiral-odd and T-odd Collins function 
$H_1^\perp$~\cite{Collins:1993kk}.

The nonperturbative mechanism encoded in $H_1^\perp$, commonly known as Collins effect, is
based on the correlation between the transverse polarization ${\bf S}_\st$ of the 
fragmenting quark and the orientation of the hadron production plane via the mixed 
product ${\bf S}_\st \cdot {\bf k} \times {\bf P}_h$, where ${\bf k}$ and ${\bf P}_h$ 
are the quark and hadron 3-momenta, respectively. 
Because of residual interactions occurring inside the jet, 
the T-odd Collins function is sensitive to the phase difference originating from the 
interference of different production channels. As a consequence, 
the transverse polarization of the quark influences the azimuthal distribution of the 
detected pions, producing the observed SSA. Without such residual interactions, the T-odd
$H_1^\perp$ vanishes and the azimuthal distribution of hadrons is perfectly symmetric. But
also an intrinsic transverse component of ${\bf k}$ is necessary for the Collins
effect to survive. In fact, in collinear approximation (i.e., assuming  ${\bf k}$ and 
${\bf P}_h$ to be parallel) the mixed product vanishes and so does the related SSA. 

The need of including the relative transverse momentum between quarks and hadrons brings 
about several complications: it is much harder to provide factorization 
proofs~\cite{Ji:2004xq}, to verify the universality of the functions 
involved~\cite{Collins:2002kn,Metz:2002iz,Boer:2003cm,Bomhof:2004aw,Collins:2004}, 
and to study their evolution
equations~\cite{Boer:2001he,Henneman:2001ev,Kundu:2001pk,Idilbi:2004vb}. Moreover, for the 
considered SSA in proton-proton collisions the transverse-momentum dependent elementary
cross section gets a suppression factor as an inverse power of the hard scale of the 
process~\cite{Anselmino:1998yz,Anselmino:2004ky}. Finally, the rich structure of the
cross section allows for other competing mechanisms leading to the same SSA, such as the 
Sivers effect~\cite{Sivers:1990cc}, the Qiu-Sterman effect~\cite{Qiu:1991pp}, and maybe 
more~\cite{Boer:1999mm}. A lively discussion is ongoing in this field in order to 
correctly interpret the available data for both hadronic collisions and semi-inclusive DIS 
(see, for example, 
Refs.~\cite{Efremov:2003tf,Schweitzer:2003yr,Bacchetta:2004zf,D'Alesio:2004up,Anselmino:2004ky,Ma:2004tr}). 

It has been pointed out already that selecting a more exclusive 
channel in the final state, where two unpolarized hadrons are detected inside the 
same jet, is obviously more challenging from the experimental point of view, but it 
represents a more convenient theoretical situation~\cite{Bacchetta:2003vn}.\footnote{Theoretical studies on 
factorization and evolution of dihadron fragmentation functions have recently 
appeared~\cite{deFlorian:2003cg,Majumder:2004wh}.}  
In particular, two vectors are available: the 
center-of-mass (cm) momentum of the pair, $P_h=P_1+P_2$, and its relative momentum 
$R=(P_1-P_2)/2$. Therefore, even after integrating the fragmentation functions upon 
${\bf P}_{h\st}$, the intrinsic transverse component of $P_h$ with respect to the jet axis, 
it is still possible to relate the transverse polarization of the fragmenting quark to 
the transverse component of the relative momentum, ${\bf R}_{\st}$, via the mixed product 
${\bf S}_{\st} \cdot {\bf R} \times {\bf P}_h$~\cite{Collins:1994ax,Jaffe:1998hf,Bianconi:1999cd}. 
For the lepton-induced production of two unpolarized hadrons (e.g., two pions) in semi-inclusive 
DIS, the generated SSA at leading twist is free from the problems mentioned above about 
the Collins effect, since the center of mass of the hadron pair is traveling collinear with 
the jet axis and, thus, collinear factorization is preserved. Indeed, the transversity $h_1$ 
can be coupled to a new class of (chiral-odd and T-odd) fragmentation functions, the so-called 
Interference Fragmentation Functions (IFF)~\cite{Radici:2001na,Bacchetta:2002ux}, that can be 
extracted independently in the corresponding process 
$e^+e^- \to (h_1 h_2)_{jet_C} (h_1 h_2)_{jet_D}
X$~\cite{Boer:2003ya,Artru:1996zu}, 
where now two pairs of 
leading hadrons are detected in each back-to-back jet. Measurements of IFF in semi-inclusive 
DIS and $e^+e^-$ are under way by the HERMES and BELLE collaborations,
respectively, and could be performed by the COMPASS and BABAR collaborations, too.

In this paper, we extend the study of IFF and inclusive two-hadron production to the 
proton-proton collision case (see also Ref.~\cite{Tang:1998wp}). 
In Sect.~\ref{sec:pair}, we describe the process $pp^\uparrow \to (h_1 h_2)_{jet}X$,
where one proton is transversely polarized and one hadron pair is detected inside
a jet. 
We show that only one source of SSA survives at leading 
twist without any suppression of the hard elementary cross section: it involves the 
convolution $f_1\otimes h_1 \otimes H_1^{\open}$, containing the usual unpolarized 
distribution $f_1$, the transversity $h_1$ and the IFF $H_1^{\open}$. No other 
mechanism is active and the theoretical situation is very clean.  

In Sect.~\ref{sec:2pairs} we study the unpolarized collision 
$pp \to (h_1 h_2)_{jet_C} (h_1 h_2)_{jet_D} X$, where two hadron pairs in separate 
jets are detected. Two leading-twist Fourier components arise in the azimuthal
orientation of the two planes (each one containing one hadron pair) with respect to the
scattering plane. One term offers the possibility of observing for the first time 
the effect of gluon linear polarization in fragmentation processes. The other one 
is proportional to the 
convolution $f_1\otimes f_1 \otimes H_1^{\open} \otimes H_1^{\open}$; hence, it can be 
used to measure the unknown IFF $H_1^{\open}$. 

Our work demonstrates how 
in (polarized) proton-proton 
collisions it is possible in principle to determine at the same time the transversity 
distribution $h_1$ and the IFF $H_1^{\open}$, without having to resort to $e^+e^-$ 
annihilation. 
The same formalism can be applied to collisions involving (un)polarized antiprotons.
Therefore, both processes can be studied not only at experiments such as STAR and PHENIX
at RHIC, but also at the future ones being planned at the above mentioned GSI HESR. 
This one and other conclusions are expanded in Sect.~\ref{sec:end}.

\section{Production of a single pair}
\label{sec:pair}

We consider first the process $A+B\rightarrow (C_1 C_2)_C+X$, where two protons (with 
momenta $P_A, P_B$, and spin vectors $S_A, S_B$) collide, and two unpolarized hadrons 
$C_1, C_2$, are inclusively detected inside the same jet $C$. The outgoing hadrons have 
masses $M_{C1}$ and $M_{C2}$, invariant mass $M_C$ and momenta $P_{C1}$ and
$P_{C2}$. It is convenient to introduce 
the vectors $P_C=P_{C1}+P_{C2}$ and $R_C=(P_{C1}-P_{C2})/2$, the total and relative 
momenta of the pair, respectively (they correspond to the general definition of $P_h$ and 
$R$ in the Introduction). The angle $\theta_C$ is the polar angle in the pair's center of 
mass between the direction of emission (which happens to be back-to-back in this frame) 
and the direction of ${\bf P}_C$ in any other frame~\cite{Bacchetta:2002ux}. The intrinsic transverse 
component of $P_C$ with respect to the jet axis, i.e.\ ${\bf P}_{C\st}$, is integrated over 
and, consequently, ${\bf P}_C$ is taken parallel to the jet axis itself. The component of $P_C$ 
perpendicular to the beam direction (defined by ${\bf P}_A$) will be denoted as 
${\bf P}_{C\perp}$.\footnote{In the following, for any vector $V$ pertinent to a hadron 
$h$ we will use the index $T$ to mean the transverse component of $V$ with respect to 
the direction of the hadron 3-momentum ${\bf P}_h$, such that ${\bf V}_{\st}\cdot 
{\bf P}_h = 0$; $V$ can be the hadron polarization vector $S_h$ (fom which the index $T$
is sometimes used also to mean the transverse polarization), the momentum $p$ of a 
parton inside $h$, etc. With the index $\perp$ we mean the transverse component of $V$ 
with respect to the incident beam direction identified by ${\bf P}_A$.}
Its modulus will serve as the hard scale of the process and it is assumed to be much bigger 
than the masses of the colliding hadrons and of $M_C$. Our analysis is valid
only at leading order in
$1/|{\bf P}_{C\perp}|$, i.e.\ at leading twist. 

The cross section for this process can be written, in analogy with single hadron 
production~\cite{Anselmino:1999pw,Anselmino:2004ky}, as
\begin{multline}
\frac{\de \sigma}{\de \eta_C\, \de |{\bf P}_{C\perp}|\, \de \cos {\theta_C}\,
  \de M_C^2\, \de \phi_{R_C}\de \phi_{S_A} \de \phi_{S_B} } 
= 2 \, |{\bf P}_{C\perp}| \,
\sum_{a,b,c,d}\frac{1}{4}\sum_{({\rm all}\; \chi{\rm's}) 
}
\int \frac{\de x_a \de x_b \de z_c}{4 \pi^2 z_c^2} \, \Phi'_a(x_a,S_A)_{\chi_a'
  \chi^{}_a} \, \Phi'_b(x_b,S_B)_{\chi_b' \chi^{}_b} \\
\frac{1}{16 \pi \hat{s}^2} \, \hat{M}_{\chi^{}_c, \chi^{}_d; \chi^{}_a,  \chi^{}_b} \, 
    \hat{M}^{\ast}_{\chi_a', \chi_b'; \chi_c' \chi_d'} \, 
    \Delta'_c (z_c,\cos{\theta_C},M_C^2,\phi_{R_C})_{\chi_c' \chi^{}_c}  \, 
    \delta_{\chi_d' \chi^{}_d} \, \hat{s} \, \delta(\hat{s}+\hat{t}+\hat{u}) \; ,
\label{eq:1pairxsect0}
\end{multline}
where $\eta_C$ is the pseudorapidity of the hadron pair, defined with respect
to ${\bf P}_A$.

\begin{figure}
\centering
\includegraphics[width=9cm]{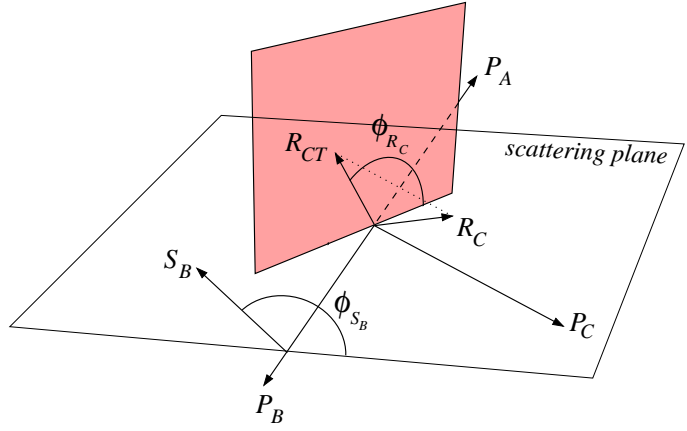}
\caption{Description of the kinematics for the proton-proton annihilation into a single pair 
in a jet.}
\label{fig:ppplanes}
\end{figure}

The azimuthal angles are defined in the hadronic center of mass as follows 
(see also Fig.~\ref{fig:ppplanes})
\begin{align} 
\cos \phi_{S_A} &= 
  \frac{(\hat{\bf P}_A \times {\bf P}_C)}{|\hat{\bf P}_A\times{\bf P}_C|}
  \cdot \frac{(\hat{\bf P}_A\times{\bf S}_A)}{|\hat{\bf P}_A
     \times{\bf S}_A|},
&
\sin \phi_{S_A} &= 
  \frac{({\bf P}_C \times {\bf S}_A) \cdot \hat{\bf P}_A}{|\hat{\bf P}_A
     \times{\bf P}_C|\,|\hat{\bf P}_A\times{\bf S}_A|} , \\
\label{eq:azang2}
\cos \phi_{S_B} &= 
  \frac{(\hat{\bf P}_B \times {\bf P}_C)}{|\hat{\bf P}_B\times{\bf P}_C|}
  \cdot \frac{(\hat{\bf P}_B\times{\bf S}_B)}{|\hat{\bf P}_B
     \times{\bf S}_B|},
&
\sin \phi_{S_B} &= 
  \frac{({\bf P}_C \times {\bf S}_B) \cdot \hat{\bf P}_B}{|\hat{\bf P}_B
     \times{\bf P}_C|\,|\hat{\bf P}_B\times{\bf S}_B|} , \\
\label{eq:azang3}
\cos \phi_{R_C} &= 
  \frac{(\hat{\bf P}_C \times {\bf P}_A)}{|\hat{\bf P}_C\times{\bf P}_A|}
  \cdot \frac{(\hat{\bf P}_C\times{\bf R}_C)}{|\hat{\bf P}_C
     \times{\bf R}_C|},
&
\sin \phi_{R_C} &= 
  \frac{({\bf P}_A \times {\bf R}_C) \cdot \hat{\bf P}_C}{|\hat{\bf P}_C
     \times{\bf P}_A|\,|\hat{\bf P}_C\times{\bf R}_C|}, 
\end{align}  
where $\hat{\bf P} ={\bf P}  /|{\bf P}| $.

The partons involved in the elementary scattering have momenta $p_a = x_a P_A$, $p_b = x_b 
P_B$, and $p_c = P_C/z_c$.
The indices $\chi$'s refer to the chiralities/helicities of the partons. The partonic hard 
amplitudes $\hat{M}$ can be taken from Ref.~\cite{Gastmans:1990xh} and are written in 
terms of the partonic Mandelstam variables $\hat{s},\;\hat{t},\;\hat{u}$, which are 
related to the external ones by 
\begin{align}
  \label{eq:mandelstam}
  \hat{s} &= x_a x_b\, s, & \hat{t} &= \frac{x_a}{z_c}\,t, & \hat{u}=  \frac{x_b}{z_c}\,u \; .
\end{align} 
Conservation of momentum at the partonic level implies that
\begin{equation} 
\label{eq:stu}
\begin{split}
  \hat{s}\, \delta(\hat{s}+\hat{t}+\hat{u}) &= z_c\, \delta\left(z_c + \frac{x_a t
    + x_b u}{x_a x_b s}\right) 
\,\stackrel{\mbox{\tiny c.m.s.}}{=}\,
z_c\, \delta\left(z_c-\rrho \frac{x_a e^{-\eta_C}
    + x_b e^{\eta_C}}{x_a x_b}\right) \, \equiv \, z_c \, \delta \left( z_c - \bar{z}_c
    \right) \; . 
\end{split}
\end{equation}
Therefore, Eq.~(\ref{eq:1pairxsect0}) can be written as
\begin{equation}
\begin{split}
\frac{\de \sigma}{\de \eta_C\, \de |{\bf P}_{C\perp}|\, \de \cos {\theta_C}\,
  \de M_C^2\, \de \phi_{R_C}\, \de \phi_{S_A} \,\de \phi_{S_B} } & = 2 \, |{\bf P}_{C\perp}|
\sum_{a,b,c,d}\frac{1}{4}\sum_{({\rm all}\; \chi{\rm's}) 
}
\int \frac{\de x_a \de x_b}{4 \pi^2 z_c} \, \Phi'_a(x_a,S_A)_{\chi_a'
  \chi^{}_a} \, \Phi'_b(x_b,S_B)_{\chi_b' \chi^{}_b} \\
 & \quad\frac{1}{16 \pi \hat{s}^2} \, \hat{M}_{\chi^{}_c, \chi^{}_d; \chi^{}_a,  \chi^{}_b} 
 \, \hat{M}^{\ast}_{\chi_a', \chi_b'; \chi_c' \chi_d'} \, 
    \Delta'_c (\bar{z}_c,\cos{\theta_C},M_C^2,\phi_{R_C})_{\chi_c' \chi^{}_c}  \, 
    \delta_{\chi_d' \chi^{}_d}\;,
\end{split}
\label{eq:1pairxsect}
\end{equation}
where $z_c$ is fixed to $\bar{z}_c$ by Eq.~(\ref{eq:stu}).

The building blocks of Eq.~(\ref{eq:1pairxsect}) are the following. The correlators $\Phi'$
describe the distribution, inside each parent hadron $A$ or $B$, of the two partons entering 
the elementary vertex and carrying a fraction $x_a$ or $x_b$, respectively, of the hadron 
momentum. Parton and corresponding hadron momenta are taken to be parallel, since intrinsic relative 
transverse components are integrated over. When the partons are quarks, then we have at leading 
twist and in the quark helicity basis~\cite{Jaffe:1992ra,Bacchetta:1999kz}
\begin{equation}
  \Phi'_a(x_a,S_A)_{\chi_a' \chi^{}_a} = \left( 
    \begin{array}{cc}
    f_1(x_a) + S_{A\, L} \, g_1(x_a) & |{\bf S}_{A\, T}| \, e^{-i\phi_{S_A}} \, h_1(x_a) 
    \\[5pt]
     |{\bf S}_{A\, T}| \, e^{i\phi_{S_A}} \, h_1(x_a) & f_1(x_a) - S_{A\, L} \, g_1(x_a) 
     \end{array} 
     \right) \; ,
\label{eq:phia}
\end{equation}
where we omitted flavor indices; $f_1$ and $g_1$ are the unpolarized and helicity 
distributions of quark $a$ in proton $A$. The correlator is written in the helicity basis
where ${\bf P}_A$ defines the $\hat{z}$ axis, with $S_{A\, L}$ and $|{\bf S}_{A\, T}|$ 
indicating the parallel and transverse components of the polarization ${\bf S}_A$ with 
respect to ${\bf P}_A$, and the $\hat{x}$ axis is oriented along ${\bf P}_{C\perp}$. 
For quark $b$ in hadron $B$ we have a similar correlator by replacing $a,\,A$  with $b,\,B$; 
consequently, the $\hat{z}$ axis is now pointing along ${\bf P}_B$ and the helicity 
$S_{B\, L}$ is considered positive or negative with respect to this axis. 
        
When the partons are gluons, the correlator reads~\cite{Jaffe:1996ik,Mulders:2000sh}
\begin{equation}
  \Phi'(x_a,S_A)_{\chi_a' \chi^{}_a} = \left( 
    \begin{array}{cc}
    G(x_a) + S_{A\, L} \, \Delta \, G (x_a) & 0 \\[5pt]
    0 & G(x_a) - S_{A\, L} \, \Delta G (x_a) \end{array} \right) \; ,
\label{eq:gluon}
\end{equation}
where $G(x), \Delta G(x)$, are the unpolarized and helicity gluon distributions; an analogous
formula holds for gluon $b$ in hadron $B$.
To our purposes, it is very important to note that in a spin-$\half$ target there is 
no gluon transversity because of the mismatch in the helicity change~\cite{Jaffe:1996zw}; 
hence, the off-diagonal elements in Eq.~(\ref{eq:gluon}) are vanishing.

The last ingredient in Eq.~(\ref{eq:1pairxsect}) is the correlator $\Delta'_c$ describing the 
fragmentation of the parton $c$ into two hadrons. 
Up to leading twist, the correlator for a fragmenting quark $c$ 
reads~\cite{Bacchetta:2002ux}
\begin{equation}
  \Delta'_c(z_c,\cos{\theta_C},M_C^2,\phi_{R_C})_{\chi_c' 
  \chi_c^{}} = \frac{1}{4 \pi} \, \left( \begin{array}{cc}
       D_1^c & i e^{i\phi_{R_C}} \, \frac{|{\bf R}_{C}|}{M_C} \, 
       \sin{\theta_C} \, H_1^{\open c} \\[5pt]
       -i e^{-i\phi_{R_C}} \, \frac{|{\bf R}_{C}|}{M_C}\, \sin{\theta_C} \, H_1^{\open c} & 
       D_1^c \end{array} \right) \; ,
\label{eq:delta}
\end{equation}
where the fragmentation functions inside the matrix depend on $z_c,\,\cos{\theta_C},\,M_C^2,$ and
\begin{equation}
|{\bf R}_C| = \frac{1}{2}\sqrt{M_C^2 - 2(M_1^2+M_2^2) + (M_1^2-M_2^2)^2/M_C^2} \; .
\label{eq:rmod}
\end{equation}
The correlator is written in the helicity basis by choosing this time the $\hat{z}$ axis along 
${\bf P}_C$ and the $\hat{x}$ axis to point along the component of ${\bf P}_A$ orthogonal to 
${\bf P}_C$ (see Fig.~\ref{fig:ppplanes}).

The fragmentation functions 
inside the correlator~(\ref{eq:delta}) can be expanded in relative partial 
waves using the basis of Legendre polinomials through the $\cos \theta_C$ 
dependence~\cite{Bacchetta:2002ux}; truncating the expansion at $L=1$, we have
\begin{equation} 
\begin{split} 
D_1^{}(z_c,\cos{\theta_C},M_C^2) &\to D_{1,oo}^{}(z_c,M_C^2) + 
D_{1,ol}^{}(z_c,M_C^2)\, \cos\theta_C + D_{1,ll}^{}(z_c,M_C^2) \, \frac{1}{4}\,
(3\cos^2\theta_C -1) \; , \\
\sin \theta_C \, H_1^{\open}(z_c,\cos{\theta_C},M_C^2) &\to  H_{1,ot}^{\open}(z_c,M_C^2) \, 
\sin \theta_C + H_{1,lt}^{\open}(z_c,M_C^2) \, \sin \theta_C \, \cos\theta_C \; . 
\end{split}
\label{eq:twist2pw}
\end{equation}
Since the fragmentation functions at leading twist are probability densities, 
the double-index notation refers to 
the polarization state in the center of mass of the pair for each separate probability 
amplitude; namely, $D_{1,oo}$ describes the decay probability into two hadrons in a relative 
$L=0$ wave, while $D_{1,ol}$ describes the interference between the amplitudes for the decay 
into a $L=0$ pair and a $L=1$ pair "longitudinally" polarized along ${\bf P}_C$ in its 
cm frame (hence, proportional to $\cos \theta_C$). Similarly, $H^{\open}_{1,ot}$ 
describes the interference between a $L=0$ pair and a "transversely" polarized $L=1$ pair 
(proportional to $\sin \theta_C$), while $D_{1,ll}$ and $H^{\open}_{1,lt}$ refer to 
interferences between different polarization components of $L=1$ pairs. 

In general, the invariant-mass dependence of the fragmentation functions is
unknown. However, we know that the $p$-wave contribution originates
essentially from a spin-1 resonance (e.g., the $\rho$ meson in two-pion
production).\footnote{The $p$-wave contributions are identical to vector-meson
fragmentation functions, which have been studied in several articles in the
context of different experiments~\cite{Efremov:1982sh,Anselmino:1984af,Ji:1994vw,Anselmino:1996vq,Anselmino:1999cg,Bacchetta:2000jk,Xu:2001hz,Xu:2003fq}.} 
Therefore, we can expect the pure $p$-wave fragmentation
functions, $D_{1,ll}$ and $H^{\open}_{1,lt}$, 
to display the invariant mass dependence typical of the resonance
(e.g., a Breit-Wigner peaked at the $\rho$ mass). The fragmentation function 
$D_{1,oo}$ contains both $s$ and $p$-wave contributions, and should therefore
look as the superposition of the resonance peak and of a continuum background.
For what concerns the
interference terms, $D_{1,ol}$ and $H^{\open}_{1,ot}$, since both $s$
and $p$ channels have to be present, they should be sizeable
only in the neighborhood of the resonance, possibly with a sign change at the position of the 
resonance peak~\cite{Jaffe:1998hf}.

When parton $c$ is a gluon, the correlator $\Delta'_c$ describes the yet unexplored
fragmentation of a gluon into two hadrons. Due to angular momentum conservation, we know that 
for gluons the off-diagonal elements of $\Delta'_c$ can contain 
only pure $L=1$ contributions from the fragmentation
of a gluon into a spin-1 resonance~\cite{Anselmino:1984af}. The
correlator at leading twist can be obtained in analogy to the case of gluon distributions in
spin-1 targets~\cite{Jaffe:1989xy,Schafer:1999am}:
\begin{equation}
  \Delta'_c (z_c,\cos{\theta_C},M_C^2,\phi_{R_C})_{\chi_c' 
  \chi_c^{}} = \frac{1}{4 \pi} \, \left( \begin{array}{cc}
       \hat{G}^c & i e^{2i\phi_{R_C}}\, \frac{|{\bf R}_{C}|^2}{M_C^2}\,
       \sin^2{\theta_C} \, \Gangle[c]\\[5pt]
       -i e^{-2i\phi_{R_C}}\, \frac{|{\bf R}_{C}|^2}{M_C^2} \, \sin^2{\theta_C} \, 
       \Gangle[c] & \hat{G}^c  \end{array} \right) \; .
\label{eq:deltagluon}
\end{equation}
The function $\hat{G}$ describes the decay of an unpolarized gluon into two unpolarized 
hadrons. Its partial-wave expansion is the same as that of $D_1$
in Eq.~(\ref{eq:twist2pw}). 
The function $\Gangle[]$ describes the decay into two unpolarized 
hadrons of a transversely polarized gluon.\footnote{The 
``transverse'' polarization of a gluon can be defined in strict analogy to the quark case. 
Indicating the positive/negative helicity states along a certain $\hat{z}$ axis by $|\pm
\rangle$, we introduce the states $|\uparrow\rangle = 1/\sqrt{2}\, (|+\rangle +  
|-\rangle)$ and $|\downarrow\rangle = i/\sqrt{2}\, (|+\rangle - |-\rangle)$. They 
correspond to {\em linearly} polarized gluons along the two independent directions 
orthogonal to $\hat{z}$, i.e.\ the $\hat{x}$ and $\hat{y}$ axis~\cite{Artru:1990zv}. 
However, to keep a uniform notation and to avoid confusion 
between linear and longitudinal polarization, we prefer to talk about transverse polarization 
states and to use the superscript $\uparrow$ when necessary (note that a
different notation was used, e.g., in Ref.~\cite{Vogelsang:1998yd}).}  
We choose the 
symbol $\Gangle[]$ for the new function to indicate a transverse gluon fragmentation 
function that needs an explicit dependence upon the relative transverse momentum 
between the two hadrons. Unfortunately, $\Gangle[]$ cannot 
appear in connection with the quark transversity distribution $h_1$ because of the mismatch 
in the units of helicity flip between a spin-$\half$ and a spin-1 objects, leading to the 
$\exp(i\phi_{R_C})$ and $\exp(2i\phi_{R_C})$ dependences in Eq.~(\ref{eq:phia}) and 
Eq.~(\ref{eq:deltagluon}), respectively. 
It can be coupled to the gluon transversity $\delta G$, which is defined only
in targets with spin 
greater than $\half$~\cite{Jaffe:1989xy,Artru:1990zv}. 
If we truncate the partial-wave expansion to $L=1$, due to angular-momentum
conservation 
$\Gangle[]$ contains only the $L=1$ contribution and reduces to the
fragmentation of a gluon into a vector meson~\cite{Anselmino:1984af}. Consequently, it is
possible to predict that it will follow the invariant-mass shape
of a spin-1 resonance. The $\theta_C$
dependence is given only by the $\sin^2 \theta_C$ prefactor, typical of a
transversely polarized resonance.

When in the proton-proton collision one of the two protons is transversely polarized, 
namely the process $pp^\uparrow \to (h_1 h_2)_{jet} X$, the most interesting SSA is 
\begin{equation}
A_N (\eta_C, |{\bf P}_{C\perp}|, \cos {\theta_C}, M_C^2, \phi_{R_C}, \phi_{S_B})
= \frac{\de \sigma^\uparrow - \de \sigma^\downarrow}
{\de \sigma^\uparrow + \de \sigma^\downarrow } \equiv 
\frac{\de \sigma_{UT}}{\de \sigma_{UU}} \; ,
\label{eq:AN}
\end{equation}
where the proton transversity distribution $h_1$ can be observed at leading twist in 
connection with the IFF $H_1^{\open}$. The longitudinal spin asymmetry displays no new 
features compared to the case where only one hadron is produced.

For the unpolarized cross section in the denominator of Eq.~(\ref{eq:AN}) (integrated over 
$\phi_{S_A}$), we have
\begin{equation}
\de \sigma_{UU}=  2 \, |{\bf P}_{C\perp}| \, \sum_{a,b,c,d}\int \frac{\de x_a
  \de x_b }{8 \pi^2 z_c} \, f_1^a(x_a) 
\, f_1^b(x_b) \, \frac{\de \hat{\sigma}_{ab \to cd}}{\de \hat{t}} \, 
D_1^c(\bar{z}_c,\cos{\theta_C},M_C^2) \; .
\label{eq:sigmaOO}
\end{equation}
Here and in the following expressions, it is understood that when the parton is a gluon, we 
need to make the replacements $f_1 \to G$ and $D_1 \to \hat{G}$. The 
unpolarized elementary cross sections are well known~\cite{Gastmans:1990xh}. For convenience, 
we rewrite them in Eqs.~(\ref{eq:elemOO-1}-\ref{eq:elemOO-4}) in the Appendix.

For the transversely polarized cross section in the numerator of
Eq.~(\ref{eq:AN}) (integrated over $\phi_{S_A}$), we have
\begin{equation}
\begin{split}
\de \sigma_{UT}&=2 \, |{\bf P}_{C\perp}|\, \sum_{a,b,c,d}\, |{\bf S}_{B\st}| \, 
\sin{(\phi_{S_B}-\phi_{R_C})} \int \frac{\de x_a \de x_b }{8 \pi^2 z_c} \, f_1^a(x_a) \, 
h_1^b(x_b) \, \frac{\de \Delta \hat{\sigma}_{ab^\uparrow \to c^\uparrow d}}{\de \hat{t}} \\
& \quad \times \frac{|{\bf R}_{C}|}{M_C}\, \sin{\theta_C} \, H_1^{\open c}(\bar{z}_c,
\cos{\theta_C}, M_C^2) \\
&\approx 2 \, |{\bf P}_{C\perp}|\, \sum_{a,b,c,d}\, \frac{|{\bf R}_{C}|}{M_C}\,
|{\bf S}_{B\st}|\, \sin{(\phi_{S_B}-\phi_{R_C})} \int \frac{\de x_a \de x_b  }{8 \pi^2 z_c} \, 
f_1^a(x_a) \, h_1^b(x_b) \, \frac{\de \Delta \hat{\sigma}_{ab^\uparrow \to c^\uparrow d}}
{\de \hat{t}} \\
 & \quad \times\sin{\theta_C}\left( H_{1,ot}^{\open c}(\bar{z}_c,M_C^2)+\cos{\theta_C} \, 
 H_{1,lt}^{\open c}(\bar{z}_c,M_C^2)\right)
\; .
\label{eq:sigmaOT}
\end{split}
\end{equation}
The elementary cross sections with transversely polarized partons $b$ and $c$ correspond to
\begin{equation}
\frac{\de \Delta \hat{\sigma}_{ab^\uparrow \to c^\uparrow d}}{\de \hat{t}} \equiv 
\frac{1}{16 \pi \hat{s}^2}\,\frac{1}{4}\sum_{({\rm all}\; \chi{\rm's})} \, 
\hat{M}_{\chi^{}_c, \chi^{}_d; \chi^{}_a, \chi^{}_b} \,  
\hat{M}^{\ast}_{\chi_a, -\chi_b; -\chi_c, \chi_d}.
\end{equation} 
They describe the cross section for the case when quark $b$ is transversely polarized in a 
direction forming an azimuthal angle $\phi_{S_b}$ around ${\bf P}_B$ and the transverse 
polarization of quark $c$ forms the same azimuthal angle $\phi_{S_c}=\phi_{S_b}$ around 
${\bf P}_C$ ($\phi_{S_b}$ and $\phi_{S_c}$ are defined analogously to $\phi_{S_B}$ and 
$\phi_{R_C}$, respectively -- see Eqs.~(\ref{eq:azang2}) and (\ref{eq:azang3}) and 
Fig.~\ref{fig:ppplanes}). We list them explicitly in 
Eqs.~(\ref{eq:elemOTini}-\ref{eq:elemOT}) in the Appendix (see also 
Ref.~\cite{Stratmann:1992gu}).

It is possible to integrate the cross sections over $\cos \theta_C$ to obtain
\begin{align}
 \de \sigma_{UU}&=  2 \, |{\bf P}_{C\perp}| \, \sum_{a,b,c,d}\int \frac{\de x_a
  \de x_b }{4 \pi^2 z_c} \, f_1^a(x_a) 
\, f_1^b(x_b) \, \frac{\de \hat{\sigma}_{ab \to cd}}{\de \hat{t}} \, 
D_{1,oo}^{}(\bar{z}_c,M_C^2), \\
\de \sigma_{UT}&= 2 \, |{\bf P}_{C\perp}|\, \sum_{a,b,c,d}\, 
\frac{|{\bf R}_{C}|}{M_C}\,|{\bf S}_{B\st}|\, \sin{(\phi_{S_B}-\phi_{R_C})} \int 
\frac{\de x_a \de x_b  }{16 \pi z_c} \, f_1^a(x_a) \, h_1^b(x_b) \, 
\frac{\de \Delta \hat{\sigma}_{ab^\uparrow \to c^\uparrow d}}{\de \hat{t}} 
  H_{1,ot}^{\open c}(\bar{z}_c,M_C^2)
\; .
\label{eq:sigmaOT2}
\end{align} 
Eqs.~(\ref{eq:sigmaOT}) and (\ref{eq:sigmaOT2}) are the most relevant results of this Section 
and a few comments are in order. First of all, we obtain a formula reminiscent of the 
original one proposed by Tang, Eq.~(8) in Ref.~\cite{Tang:1998wp}. However, there are some 
crucial differences: in Ref.~\cite{Tang:1998wp} the dependence on the azimuthal angles is 
different, the connection to the external variables (rapidity and transverse momentum of the
pair) is not made clear, and the behavior in the invariant mass is factorized out of the 
IFF, which is a model-dependent statement. Finally, a couple of differences in the elementary 
cross sections are pointed out in the Appendix. Our asymmetries are also analogous to the 
ones for the process $pp^{\uparrow} \to \Lambda^{\uparrow} X$ discussed in 
Refs.~\cite{deFlorian:1998ba,deFlorian:1998am}. In that case, however, the asymmetry is 
proportional to $\cos{(\phi_{S_B}-\phi_{S_\Lambda})}$ (with $\phi_{S_\Lambda}$ being the
azimuthal angle of the transverse spin of the $\Lambda$, defined analogously
to $\phi_{R_C}$): the transverse spin of the quark is directly transferred to the transverse 
spin of the $\Lambda$, 
while here it is connected to ${\bf R}_{C\, \st}$ via the mixed product entailed in the T-odd
$H_1^{\open}$, which implies an extra $\textstyle{\frac{\pi}{2}}$ rotation. 
Finally, our asymmetry can be related to that occurring in two-hadron
production in DIS~\cite{Radici:2001na,Bacchetta:2002ux}, by replacing $f_1(x_a)$ with 
$\delta(1-x_a)$, using the elementary cross section for $l q^\uparrow \to q^\uparrow l$ and 
taking into account the fact that the final parton $d$ is also observed, so that $x_b$ and 
$z_c$ are fixed according to Eq.~(\ref{eq:conservation}) of next Sect.~\ref{sec:2pairs} with 
$z_d=1$. In Refs.~\cite{Radici:2001na,Bacchetta:2002ux}, the asymmetry depends on 
$\sin{(\phi_{S_B}+\phi_{R_C})}$ simply because the azimuthal angles are defined in a 
different way with respect to the present work.

Quantitative estimates of these asymmetries are possible either by using models for the 
IFF~\cite{Jaffe:1998hf,Bianconi:1999uc,Radici:2001na} or by saturating their positivity 
bounds~\cite{Bacchetta:2002ux}. Measurements of IFF are under way at HERMES (semi-inclusive 
DIS) and at BELLE~\cite{GrossePerdekamp:2002eb} ($e^+e^-$ annihilation). Experimental results 
at different energies can be related through the same DGLAP equations applicable to the
fragmentation function $H_1$~\cite{Stratmann:2001pt}.\footnote{Note that 
we deal with
functions that depend explicitly on the (limited) relative transverse momentum of
the hadron pair.  
On the contrary, the
  evolution equations studied in Refs.~\cite{deFlorian:2003cg,Majumder:2004wh}
apply to dihadron fragmentation functions integrated over the relative
transverse momentum of the hadron pair.} However, as we shall see in the next 
Section, IFF can be measured independently in the very same proton-proton collisions analyzed 
so far.  

\section{Production of two pairs}
\label{sec:2pairs}

As already anticipated in Sect.~\ref{sec:intro}, the simultaneous detection of two hadron 
pairs belonging to two separate jets allows the extraction of the specific IFF, 
$H_1^{\open}$, that occurs coupled to the transversity $h_1$ in the corresponding 
production of a single hadron pair in one jet (see previous Sect.~\ref{sec:pair}). 

The definition of the momenta and angles of the second hadron pair is done in complete 
analogy to the first pair, just replacing all indices $c,\,C$ with $d,\,D$. The correlation 
function $\Delta'_d$ is obtained in the same way. The generic expression for the cross 
section is
\begin{equation}
\begin{split}
\lefteqn{\frac{\de \sigma}{\de \eta_C\, \de |{\bf P}_{C\perp}|\, \de \cos {\theta_C}\,
  \de M_C^2\, \de \phi_{R_C}\,\de \eta_D\, \de |{\bf P}_{D\perp}|\, \de \cos {\theta_D}\,
  \de M_D^2\, \de \phi_{R_D} \de \phi_{S_A} \de \phi_{S_B}}} \\ 
 &
= 2 \sum_{a,b,c,d}\frac{1}{4}\sum_{({\rm all}\; \chi{\rm's}) 
}
\int \frac{\de x_a}{4 \pi^2}  \, \Phi'_a(x_a,S_A)_{\chi_a'
  \chi^{}_a} \, \bar{x}_b\,\Phi'_b(\bar{x}_b,S_B)_{\chi_b' \chi^{}_b} 
\frac{1}{16 \pi \hat{s}^2} \, \hat{M}_{\chi^{}_c, \chi^{}_d; \chi^{}_a,  \chi^{}_b} 
 \, \hat{M}^{\ast}_{\chi_a', \chi_b'; \chi_c' \chi_d'} \, 
\\ 
&\quad
    \Delta'_c (\bar{z}_c,\cos{\theta_C},M_C^2,\phi_{R_C})_{\chi_c' \chi^{}_c}  \, 
    \Delta'_d (\bar{z}_d,\cos{\theta_D},M_D^2,\phi_{R_D})_{\chi_d' \chi^{}_d} \; ,
\end{split}
\end{equation}
with
\begin{align}
\bar{z}_c&\stackrel{\mbox{\tiny c.m.s.}}{=} \rrho\frac{e^{\eta_C} + e^{\eta_D}}{x_a}, 
& \bar{z}_d&\stackrel{\mbox{\tiny c.m.s.}}{=}\frac{|{\bf{P}}_{D\perp}|}{\sqrt{s}} 
\frac{e^{\eta_C} + e^{\eta_D}}{x_a},
&\bar{x}_b&\stackrel{\mbox{\tiny c.m.s.}}{=} x_a e^{-\eta_C} e^{-\eta_D}.
\label{eq:conservation}
\end{align}
The above relations are obtained from momentum conservation at the partonic level. 

For unpolarized proton-proton collisions, the main observable 
is the unpolarized cross section (integrated over the angles $\phi_{S_A}$ and  $\phi_{S_B}$)
\begin{equation}
\de \sigma_{UU}  = {\cal{A}} + \cos{(\phi_{R_C}-\phi_{R_D})} \; {\cal{B}} +  
\cos{(2\phi_{R_C}-2\phi_{R_D})} \; {\cal{C}} \; .
\label{eq:2pairxsec}
\end{equation}
The function ${\cal A}$ is given by 
\begin{align} 
\begin{split}
{\cal{A}} & =
\sum_{a,b,c,d}\int \frac{\de x_a}{8 \pi^2} \, f_1^a(x_a) \, \bar{x}_b\,f_1^b(\bar{x}_b) \, 
\frac{\de \hat{\sigma}_{ab \to cd}}{\de \hat{t}} \, D_1^c(\bar{z}_c,\cos{\theta_C},M_C^2) 
\, D_1^d(\bar{z}_d,\cos{\theta_D},M_D^2) \, 
\\
& \approx\sum_{a,b,c,d}\int \frac{\de x_a}{8 \pi^2} \, f_1^a(x_a) \, \bar{x}_b\,
f_1^b(\bar{x}_b) \, \frac{\de \hat{\sigma}_{ab \to cd}}{\de \hat{t}} \,  \\
& \quad \times \Bigl(D_{1,oo}^{c}(\bar{z}_c,M_C^2) + D_{1,ol}^{c}(\bar{z}_c,M_C^2)
\cos\theta_C + D_{1,ll}^{c}(\bar{z}_c,M_C^2) \frac{1}{4}\,(3\cos^2\theta_C -1)\Bigr) \\
& \quad \times \Bigl(D_{1,oo}^{d}(\bar{z}_d,M_D^2) + D_{1,ol}^{d}(\bar{z}_d,M_D^2)
\cos\theta_D + D_{1,ll}^{d}(\bar{z}_d,M_D^2) \frac{1}{4}\,(3\cos^2\theta_D -1)\Bigr) \; ,
\end{split}
\label{eq:a}
\end{align}
where the elementary cross sections $d\hat{\sigma}_{ab \to cd}$ are given by
Eqs.~(\ref{eq:elemOO-1}-\ref{eq:elemOO-4}). In other words, ${\cal A}$ is the analogue of
Eq.~(\ref{eq:sigmaOO}) for the production of two hadron pairs in separate
jets.  When the parton is a gluon, we 
need to make the replacements $f_1 \to G$ and $D_1 \to \hat{G}$.

The function ${\cal B}$ is given by
\begin{align}
\begin{split}
{\cal{B}}  &= \sum_{a,b,c,d}\int  \frac{\de x_a}{8 \pi^2} \, 
f_1^a(x_a) \, \bar{x}_b\,f_1^b(\bar{x}_b) \, 
\frac{\de \Delta \hat{\sigma}_{ab \to c^\uparrow d^\uparrow}}{\de \hat{t}} \,  
\frac{|{\bf R}_{C}|}{M_C}\, \sin\theta_C \, H_1^{\open c}(\bar{z}_c,\cos{\theta_C},M_C^2) 
\, \frac{|{\bf R}_{D}|}{M_D} \, \sin\theta_D \, H_1^{\open d}(\bar{z}_d,\cos{\theta_D},M_D^2)
\\
& \approx\sum_{a,b,c,d} \int \frac{\de x_a}{8 \pi^2} \, 
f_1^a(x_a) \, \bar{x}_b\,f_1^b(\bar{x}_b) \, 
\frac{\de \Delta \hat{\sigma}_{ab \to c^\uparrow d^\uparrow}}{\de \hat{t}} \, 
\frac{|{\bf R}_{C}|}{M_C}\, \sin\theta_C\left( H_{1,ot}^{\open c}(\bar{z}_c,M_C^2)
+\cos{\theta_C} \, H_{1,lt}^{\open c}(\bar{z}_c,M_C^2)\right)\, \\
&\quad \times \frac{|{\bf R}_{D}|}{M_D}\, \sin\theta_D 
\left(H_{1,ot}^{\open d}(\bar{z}_d,M_D^2)+\cos{\theta_D} \, 
H_{1,lt}^{\open d}(\bar{z}_d,M_D^2)\right) \; ,
\end{split}
\label{eq:b}
\end{align} 
with partons $c$ and $d$ being necessarily quarks and 
with the relevant partonic cross sections being 
\begin{equation}
 \frac{\de \Delta \hat{\sigma}_{ab \to c^\uparrow d^\uparrow}}{\de \hat{t}} \equiv  
\frac{1}{16 \pi \hat{s}^2}\,\frac{1}{4}\sum_{({\rm all}\; \chi{\rm's})} \, 
\hat{M}_{\chi^{}_c, \chi^{}_d; \chi^{}_a, \chi^{}_b} \, 
\hat{M}^{\ast}_{\chi_a, \chi_b; -\chi_c, -\chi_d} \; .
\end{equation}
They are explicitly listed in Eqs.~(\ref{eq:elemTT-1}-\ref{eq:elemTT-3}) in the Appendix.

Finally, the function ${\cal C}$ is 
\begin{equation} 
\begin{split} 
{\cal{C}}  &=  \sum_{a,b,c,d}\int  \frac{\de x_a}{8 \pi^2}\, 
f_1^a(x_a) \, \bar{x}_b\,f_1^b(\bar{x}_b) \, 
\frac{\de \Delta \hat{\sigma}_{ab \to g^\uparrow g^\uparrow}}{\de \hat{t}} \, \\
&\quad \times \frac{|{\bf R}_{C}|^2}{M_C^2}\, \sin^2\theta_C\, \Gangle[c]
(\bar{z}_c,\cos{\theta_C},M_C^2)\, \frac{|{\bf R}_{D}|^2}{M_D^2}\, \sin^2\theta_D \, 
\Gangle[d](\bar{z}_d,\cos{\theta_C},M_D^2) \\
&\approx  \sum_{a,b,c,d} \int \frac{\de x_a}{8 \pi^2}\, 
f_1^a(x_a) \, \bar{x}_b\,f_1^b(\bar{x}_b) \, 
\frac{\de \Delta \hat{\sigma}_{ab \to g^\uparrow g^\uparrow}}{\de \hat{t}} \,  \\
& \quad \times \frac{|{\bf R}_{C}|^2}{M_C^2}\, \sin^2\theta_C\, 
\Gangle[c](\bar{z}_c,M_C^2)\, \frac{|{\bf R}_{D}|^2}{M_D^2}\, \sin^2\theta_D \, 
\Gangle[d](\bar{z}_d,M_D^2) \; ,
\end{split} 
\label{eq:c}
\end{equation}  
where the nonvanishing elementary cross sections are given in Eq.~(\ref{eq:elemTT2}) in
the Appendix.

After integrating upon the angles $\theta_C$ and $\theta_D$, we obtain
\begin{equation}
 \de \sigma_{UU} = {\cal{A}}' + \cos{(\phi_{R_C}-\phi_{R_D})} \; {\cal{B}}' +
 \cos{(2\phi_{R_C}-2\phi_{R_D})}\; {\cal{C}}' \; ,
\label{eq:2pairxsecint}
\end{equation}
where
\begin{align}
\begin{split} 
{\cal{A}}' & = \int_0^\pi d\theta_C \, \sin \theta_C \int_0^\pi d\theta_D\,\sin\theta_D \, 
{\cal A} \\
&= \sum_{a,b,c,d}\int \frac{\de x_a}{2 \pi^2 } \, f_1^a(x_a) \, \bar{x}_b\,f_1^b(\bar{x}_b) 
\, \frac{\de \hat{\sigma}_{ab \to cd}}{\de \hat{t}} \,D_{1,oo}^{}(\bar{z}_c,M_C^2) \, 
D_{1,oo}^{}(\bar{z}_d,M_D^2) \; ,
\label{eq:a'} 
\end{split} 
\\
\begin{split}
{\cal{B}}' &= \int_0^\pi d\theta_C \, \sin \theta_C \int_0^\pi d\theta_D\,\sin\theta_D \, 
{\cal B} \\
&= \sum_{a,b,c,d}\int  \frac{\de x_a}{32} \, f_1^a(x_a) \, \bar{x}_b\,f_1^b(\bar{x}_b) \, 
\frac{\de \Delta \hat{\sigma}_{ab \to c^\uparrow d^\uparrow}}{\de \hat{t}} \,
\frac{|{\bf R}_{C}|}{M_C} \, H_{1,ot}^{\open c}(\bar{z}_c,M_C^2)\, \frac{|{\bf R}_{D}|}{M_D} 
\, H_{1,ot}^{\open d}(\bar{z}_d,M_D^2) \; ,
\label{eq:b'} 
\end{split}
\\
\begin{split}
{\cal{C}}' &= \int_0^\pi d\theta_C \, \sin \theta_C \int_0^\pi d\theta_D\,\sin\theta_D \, 
{\cal C} \\
&= 2 \sum_{a,b,c,d}\int  \frac{\de x_a}{9 \pi^2} \, f_1^a(x_a) \, \bar{x}_b\,f_1^b(\bar{x}_b) 
\, \frac{\de \Delta \hat{\sigma}_{ab \to g^\uparrow g^\uparrow}}{\de \hat{t}} \,
\frac{|{\bf R}_{C}|^2}{M_C^2}\, \Gangle[c](\bar{z}_c,M_C^2)\, 
\frac{|{\bf R}_{D}|^2}{M_D^2} \, \Gangle[d](\bar{z}_d,M_D^2) \; .
\label{eq:c'}
\end{split}
\end{align} 

The above expressions can be related to what has been obtained for the case of $e^+e^-$ 
annihilation~\cite{Boer:2003ya}, by replacing $f_1(x)$ with $\delta(1-x)$ and using the 
elementary cross section for $e^+e^- \to q^\uparrow \bar{q}^\uparrow$ (clearly no gluon 
contribution is present). Once again, the apparent difference in the resulting angular 
dependence is due solely to a different definition of the azimuthal angles.

Both functions ${\cal B}$ and ${\cal C}$ (or ${\cal B}'$ and ${\cal C}'$) are 
interesting. The first one contains two interference fragmentation functions
$H_1^{\open}$, one for each hadron pair: measuring the $\cos{(\phi_{R_C}-\phi_{R_D})}$
asymmetry of the cross section for the $pp \to (h_1 h_2)_{jet_C} (h_1 h_2)_{jet_D} X$
process allows the extraction of $H_1^{\open}$ and, in turn, of the transversity $h_1$ 
from the $\de \sigma_{UT}$ asymmetry described in the previous Section. The second 
observable, ${\cal C}$ (or ${\cal C}'$), describes
the fragmentation of two transversely (linearly) polarized gluons into two 
transversely (linearly) polarized spin-1 resonances.

\section{Conclusions}
\label{sec:end}

Understanding the polarization of partons inside hadrons is a fundamental goal in
order to describe the partonic structure of the hadrons themselves. At
present, the main missing piece of information is represented by the quark
transversity  distribution, $h_1$, 
a leading-twist parton density that describes the distribution of transversely polarized 
quarks inside transversely polarized hadrons. Its chiral-odd nature has prevented it from 
being measured in the simplest elementary processes like inclusive DIS. Several alternative 
strategies have been suggested in the literature, among which the most popular ones  
are transversely polarized Drell-Yan~\cite{Ralston:1979ys} and the Collins 
effect~\cite{Collins:1993kk} in semi-inclusive DIS with transversely polarized targets. 

Selecting a more exclusive final state with two detected hadrons inside the
same jet could be a more convenient option. Even when the center of mass of the
two hadrons is assumed to move collinear with the jet axis, the transverse 
component  of the relative momentum of the two hadrons (with respect to the jet axis, 
or equivalently with respect to the center-of-mass direction) 
is still available to build a single-spin asymmetry 
that singles out $h_1$ at leading twist via an 
interference fragmentation function, $H_1^{\open}$~\cite{Radici:2001na}. 
The asymmetry described by $H_1^{\open}$ is related to the azimuthal position of the 
hadron-pair plane with respect to the scattering plane. All distribution and
fragmentation functions can be integrated over intrinsic transverse momenta, 
making it simpler to deal with issues such as 
evolution equations, factorization  and universality~\cite{Collins:2004}. The comparison 
between hadron-hadron collisions, semi-inclusive DIS~\cite{Bacchetta:2003vn} and $e^+ e^-$
annihilation~\cite{Boer:2003ya} 
becomes therefore simpler than for the Collins effect.

In this paper, we have applied the formalism of interference fragmentation functions to 
proton-proton collisions. We have shown that in the production of one hadron pair in 
collisions with one transversely polarized proton, it is possible to isolate the 
convolution $f_1\otimes h_1 \otimes H_1^{\open}$, involving the usual momentum
distribution $f_1$, through the measurement of the asymmetry of the cross section 
in the azimuthal orientation of the pair around its center-of-mass momentum.
In the production of two hadron pairs in two separate jets 
in unpolarized collisions, it is possible to isolate the convolution 
$f_1 \otimes f_1 \otimes H_1^{\open}\otimes H_1^{\open}$, through the measurement of 
the asymmetry of the cross section in the 
azimuthal orientation of the two 
pairs around their center-of-mass momenta. 
Since no distribution and fragmentation functions with an explicit transverse-momentum 
dependence are required, there is no need to consider suppressed contributions
in the elementary cross sections included in the convolutions and the discussed
asymmetries remain at leading twist.
Therefore proton-proton collisions offer a unique possibility to measure
simultaneously the transversity distribution $h_1$, and the interference
 fragmentation function $H_1^{\open}$.

Finally, we have also shown that unpolarized proton-proton collisions into two hadron 
pairs (basically into two spin-1 resonances) can also provide novel information on the 
role of gluon linear polarization. 

Our formalism can be applied also to collisions involving (polarized) antiprotons.
Therefore, it can be used by experimental collaborations working on existing machines 
with polarized proton beams (like STAR and PHENIX at RHIC), but also with polarized
antiprotons at GSI HESR. Unpolarized
collisions into two hadron pairs can be studied also at Tevatron and LHC. 
In the future, we hope that it will be possible to compare 
experimental results in hadron-hadron collisions with those in semi-inclusive DIS and 
$e^+ e^-$ annihilation and perform a global analysis of dihadron
fragmentation functions.


\begin{acknowledgments}
Discussions with M.~Stratmann, B.~J\"ager are gratefully acknowledged. The work of A.~B. has 
been supported by the Alexander von Humboldt Foundation. 
\end{acknowledgments}


\appendix*
\section{Elementary cross sections}

We list here the unpolarized partonic cross sections 
\begin{equation}
\frac{\de \hat{\sigma}_{ab \to cd}}{\de \hat{t}} \equiv \frac{1}{16 \pi \hat{s}^2}\,
\frac{1}{4}\sum_{({\rm all}\; \chi{\rm's})} \, \hat{M}_{\chi^{}_c, \chi^{}_d; \chi^{}_a,  
\chi^{}_b} \, \hat{M}^{\ast}_{\chi_a, \chi_b; \chi_c \chi_d} \; ,
\end{equation} 
\begin{align}
 \frac{\de \hat{\sigma}_{qq \to qq}}{\de \hat{t}} & = 
\frac{4 \pi \alpha_s^2}{9}\left(\frac{\hat{s}^4 + \hat{t}^4 + \hat{u}^4}{\hat{s}^2 \, 
\hat{t}^2 \, \hat{u}^2} -\frac{8}{3 \, \hat{t} \, \hat{u}} \right) \; , 
&
 \frac{\de \hat{\sigma}_{qq' \to q'q}}{\de \hat{t}} & = 
\frac{4 \pi \alpha_s^2}{9}\left(\frac{\hat{s}^2 + \hat{t}^2}{\hat{s}^2 \, \hat{u}^2} \right) 
\; , \label{eq:elemOO-1} \\
 \frac{\de \hat{\sigma}_{q\bar{q} \to q'\bar{q}'}}{\de \hat{t}} & = 
\frac{4 \pi \alpha_s^2}{9}\left(\frac{\hat{t}^2 + \hat{u}^2}{\hat{s}^4} \right) \; ,
&
 \frac{\de \hat{\sigma}_{q\bar{q} \to q\bar{q}}}{\de \hat{t}} & = 
\frac{4 \pi \alpha_s^2}{9 \, \hat{s}^4 \, \hat{t}^2}\left(\hat{s}^4 + \hat{t}^4 + \hat{u}^4 -
\frac{8}{3} \, \hat{s} \, \hat{t} \, \hat{u}^2 \right) \; ,
\label{eq:elemOO-2} \\
 \frac{\de \hat{\sigma}_{q\bar{q} \to gg}}{\de \hat{t}} & = 
\frac{8 \pi \alpha_s^2}{3} \, \frac{\hat{t}^2 + \hat{u}^2}{\hat{s}^2}\left(\frac{4}{9 \, 
\hat{t} \, \hat{u}}-\frac{1}{\hat{s}^2} \right) \; ,
&
 \frac{\de \hat{\sigma}_{gq \to gq}}{\de \hat{t}} & = 
\pi  \alpha_s^2 \, \frac{\hat{s}^2 + \hat{t}^2}{\hat{s}^2}\left(\frac{1}{\hat{t}^2}- 
\frac{4}{9 \, \hat{s} \, \hat{u}} \right) \; ,
\label{eq:elemOO-3} \\
 \frac{\de \hat{\sigma}_{gg \to gg}}{\de \hat{t}} & = 
\frac{9 \pi \alpha_s^2}{8} \, \frac{(\hat{s}^4 + \hat{t}^4 + \hat{u}^4)(\hat{s}^2 + \hat{t}^2 
+ \hat{u}^2)}{\hat{s}^4 \, \hat{t}^2 \, \hat{u}^2} \; ,
&
 \frac{\de \hat{\sigma}_{gg \to q\bar{q}}}{\de \hat{t}} & =
\frac{3 \pi \alpha_s^2}{8} \, \frac{\hat{t}^2 + \hat{u}^2}{\hat{s}^2}\left(\frac{4}{9 \, 
\hat{t} \, \hat{u}}-\frac{1}{\hat{s}^2}\right) \; .
\label{eq:elemOO-4}
\end{align}
All other nonvanishing cross sections can be obtained from these ones by means of simple 
crossings.

We define the partonic cross sections differences 
with transversely polarized partons $b$ and $c$ as
\begin{equation}
\frac{\de \Delta \hat{\sigma}_{ab^\uparrow \to c^\uparrow d}}{\de \hat{t}} \equiv 
\frac{1}{16 \pi \hat{s}^2}\,\frac{1}{4}\sum_{({\rm all}\; \chi{\rm's})} \, 
\hat{M}_{\chi^{}_c, \chi^{}_d; \chi^{}_a, \chi^{}_b} \,  
\hat{M}^{\ast}_{\chi_a, -\chi_b; -\chi_c, \chi_d} \; .
\end{equation} 
In principle, these are not cross sections, but rather bilinear combinations
of amplitudes. However, they correspond to cross sections for specific polarization
states of the partons involved. 
The nonvanishing ones are
\begin{align}
\label{eq:elemOTini}
 \frac{\de \Delta \hat{\sigma}_{qq^\uparrow \to q^\uparrow q}}{\de \hat{t}} & =  
 -\frac{8 \pi \alpha_s^2}{27\hat{s}^2}\, \frac{\hat{s}\, (3\hat{t} -\hat{u}) }{\hat{u}^2} \; , 
&
\frac{\de \Delta \hat{\sigma}_{qq'^\uparrow \to q'^\uparrow q}}{\de \hat{t}} & = 
- \frac{8 \pi \alpha_s^2}{9\hat{s}^2}\, \frac{\hat{t}\,\hat{s}}{\hat{u}^2} \; ,
\\
\frac{\de \Delta \hat{\sigma}_{q\bar{q}^\uparrow \to q^\uparrow \bar{q}}}{\de \hat{t}} & = -
\frac{8 \pi \alpha_s^2}{27\,\hat{s}^2} \; , &
\frac{\de \Delta \hat{\sigma}_{q\bar{q}^\uparrow \to \bar{q}^\uparrow q}}{\de \hat{t}} & = 
-\frac{8 \pi \alpha_s^2}{27\,\hat{s}^2}\, \frac{\hat{t}\, (3\hat{s} -\hat{u}) }{\hat{u}^2} \; 
, \\
\frac{\de \Delta \hat{\sigma}_{gq^\uparrow \to q^\uparrow g}}{\de \hat{t}} & =
- \frac{8 \pi \alpha_s^2}{9\,\hat{s}^2} \left(1-\frac{9}{4}\frac{\hat{t}\,\hat{s}}{\hat{u}^2} 
\right) \; ,
&
\frac{\de \Delta \hat{\sigma}_{qg^\uparrow \to g^\uparrow q}}{\de \hat{t}} & =
- \frac{8 \pi \alpha_s^2}{9\,\hat{s}^2} \left(1-\frac{9}{4}\frac{\hat{t}\,\hat{s}}{\hat{u}^2} 
\right) \; ,
\\
\frac{\de \Delta \hat{\sigma}_{gg^\uparrow \to g^\uparrow g}}{\de \hat{t}} & =
\frac{9 \pi \alpha_s^2}{2\,\hat{s}^2} \frac{\hat{u}^2-\hat{s} \, \hat{t}}{\hat{u}^2} \; . & &
\label{eq:elemOT}
\end{align}
These cross sections correspond to the results presented in Table 1 of
Ref.~\cite{Stratmann:1992gu}, when the initial and final azimuthal angles of
the quarks (as defined in Sec.~\ref{sec:pair}) are equal (in the language of that paper, 
when  $A(s_b,s_c)=-t/2$, or equivalently when $\beta = \Phi - \pi/2$). 
They correspond also to the ``transversity dependent'' cross sections of Table I of 
Ref.~\cite{Tang:1998wp} (to compare the results, $\hat{t}$ and $\hat{u}$ have to be 
interchanged, since $a$ is the polarized parton in Ref.~\cite{Tang:1998wp}), except
for a factor 2 difference in the last term of the $q\bar{q}^\uparrow \to \bar{q}^\uparrow q$
cross section and for the absence in Ref.~\cite{Tang:1998wp} of the 
$q\bar{q}^\uparrow \to q^\uparrow \bar{q}$ cross section. The last two partonic cross 
sections are missing in both Refs.~\cite{Stratmann:1992gu} and \cite{Tang:1998wp}: they are 
less useful because there is no gluon transversity inside the
proton. We present them for completeness and for possible future applications with spin-1 
targets. As already clarified in Sec.~\ref{sec:pair}, we use the transverse gluon polarization 
states, $|\uparrow\rangle$ and $|\downarrow\rangle$, in place of linear polarization states 
along the $\hat{x}$ and $\hat{y}$ directions~\cite{Artru:1990zv} (note that a
different notation was used, e.g., in Ref.~\cite{Vogelsang:1998yd}).

We introduce the partonic cross section differences 
for two transversely polarized partons in the final
state 
\begin{equation}
 \frac{\de \Delta \hat{\sigma}_{ab \to c^\uparrow d^\uparrow}}{\de \hat{t}} \equiv 
 \frac{1}{16 \pi \hat{s}^2}\,\frac{1}{4}\sum_{({\rm all}\; \chi{\rm's})} \, 
 \hat{M}_{\chi^{}_c, \chi^{}_d; \chi^{}_a, \chi^{}_b} \, 
 \hat{M}^{\ast}_{\chi_a, \chi_b; -\chi_c, -\chi_d} \; .
\end{equation}
When the final-state partons are quarks, we have the following nonvanishing
cross sections, to be used in Eq.~(\ref{eq:b}):
\begin{align}
 \frac{\de \Delta \hat{\sigma}_{qq \to q^\uparrow q^\uparrow}}{\de \hat{t}} & =  
 -\frac{8 \pi \alpha_s^2}{27\,\hat{s}^2} \; ,
&\frac{\de \Delta \hat{\sigma}_{q\bar{q} \to q^\uparrow \bar{q}^\uparrow}}{\de \hat{t}} & = 
-\frac{8 \pi \alpha_s^2}{27\,\hat{s}^2} \, \frac{\hat{u} \, (3 \hat{t}-\hat{s}) }{\hat{s}^2} \; ,
\label{eq:elemTT-1} \\
 \frac{\de \Delta \hat{\sigma}_{q\bar{q} \to \bar{q}_T q_T}}{\de \hat{t}} & = -
 \frac{8 \pi \alpha_s^2}{27\,\hat{s}^2}\, \frac{\hat{t} \, (3 \hat{u}-\hat{s}) }{\hat{s}^2} \; ,
&
\frac{\de \Delta \hat{\sigma}_{q\bar{q} \to q'^\uparrow \bar{q}'^\uparrow}}{\de \hat{t}} & = 
-\frac{8 \pi \alpha_s^2}{9\,\hat{s}^2}\, \frac{\hat{t} \, \hat{u} }{\hat{s}^2}
\; ,
\label{eq:elemTT-2} \\
 \frac{\de \Delta \hat{\sigma}_{gg \to q^\uparrow \bar{q}^\uparrow}}{\de \hat{t}} & =
 - \frac{ \pi \alpha_s^2}{3\,\hat{s}^2} \left(1 - \frac{9}{4}\, 
 \frac{\hat{t}\, \hat{u}}{\hat{s}^2} \right) \; .
\label{eq:elemTT-3}
\end{align}
When the final-state partons are gluons, we have the following nonvanishing
cross sections, to be used in Eq.~(\ref{eq:c}):
\begin{align}
 \frac{\de \Delta \hat{\sigma}_{q\bar{q} \to g^\uparrow g^\uparrow}}{\de \hat{t}} & = 
 - \frac{64 \pi \alpha_s^2}{27\,\hat{s}^2}
 \left(1 - \frac{9}{4}\, \frac{\hat{t}\, \hat{u}}{\hat{s}^2} \right) \; ,
&\frac{\de \Delta \hat{\sigma}_{gg \to g^\uparrow g^\uparrow}}{\de \hat{t}} & =
\frac{9 \pi \alpha_s^2}{2\,\hat{s}^2}\, \frac{\hat{u}^2-\hat{s} \, \hat{t}}{\hat{s}^2} \; .
\label{eq:elemTT2}
\end{align}


\bibliographystyle{apsrev}
\bibliography{mybiblio}

\end{document}